# Extrinsic and Intrinsic Photoresponse in Monodisperse Carbon Nanotube Thin Film Transistors

Erik Sczygelski[1,∥], Vinod K. Sangwan[1,∥], Chung-Chiang Wu[1], Heather N. Arnold[1], Ken Everaerts[1,2], Tobin J. Marks[1,2], Mark C. Hersam[1,2,3], and Lincoln J. Lauhon[1,*]

[1]Department of Materials Science and Engineering, Northwestern University, Evanston, IL 60208, USA

[2]Department of Chemistry, Northwestern University, Evanston, IL 60208, USA

[3]Department of Medicine, Northwestern University, Evanston, IL 60208, USA

∥ These authors contributed equally.

* Corresponding author.

**Abstract:**
Spectroscopic, time-resolved scanning photocurrent microscopy is shown to distinguish the intrinsic photoresponse of monodisperse semiconducting (99%) single-walled carbon nanotubes (SWCNTs) from the extrinsic photoresponse of the substrate. A persistent positive photocurrent induced by near-IR excitation is attributed to the generation of free carriers by inter-band excitation in SWCNTs. For shorter excitation wavelengths, absorption by the Si substrate generates two types of photocurrent: a transient positive photoresponse, identified as a displacement current, and a persistent negative photocurrent that arises from photogating of the SWCNT thin film.

In the last decade, the unique optical properties of single-walled carbon nanotubes (SWCNTs) have provided impressive performance attributes in a wide variety of optoelectronic devices such as light-emitting diodes, photodetectors and photovoltaic cells[1,2], opto-electronic power-sources[3], and bolometric sensors[4]. The optical properties of quasi-1D SWCNTs show strong variations with diameter and chirality[5], and chirality also determines whether SWCNTs are metallic or semiconducting. Most SWCNT device research has focused on devices based on individual SWCNTs[6-9] rather than SWCNT films due to the heterogeneous mixture of chiralities in as-grown SWCNTs. However, assembly of macroscopically uniform SWCNT films may provide a more practical route towards optoelectronic applications, as the recent availability of monodisperse SWCNTs by density gradient ultra-centrifugation[10,11] techniques has enabled the realization of large-area SWCNT thin film based optoelectronic devices.[12,13] Incorporation of SWCNTs into thin film devices with other optically responsive materials, such as Si substrates, can lead to other extrinsic effects such as photogating[14] that must be differentiated from the intrinsic SWCNT response to enable rational device optimization. In addition, SWCNT thin film transistors (TFTs) show hysteresis due to trap charges in most oxide gate dielectrics that can adversely affect the intrinsic photoresponse of SWCNTs. Several recent studies of the photoresponse of SWCNTs thin films have invoked a host of different mechanisms including the bolometric effect[4,15], desorption of adsorbates[16], photothermoelectric effects[17,18], and Schottky barrier assisted separation of photogenerated carriers[8,19,20]. In this context, scanning photocurrent microscopy (SPCM) can be a useful technique for elucidating the operating principles of novel optoelectronic devices[9,21-23] and supporting the concurrent optimization of new nanostructured electronic materials.

Here, we employ a unique combination of complementary SPCM measurements to identify extrinsic and intrinsic photocurrent generation mechanisms in high purity semiconducting SWCNT TFTs. Scanning local photocurrent measurements are conducted on monodisperse SWCNT TFTs on a novel hybrid organic-inorganic high-κ self-assembled nanodielectric (SAND),[24,25] which dramatically reduces gate hysteresis and increases the transconductance relative to conventional oxide dielectrics. The small hysteresis and optimization of the ratio of channel length to width enables study of the photoresponse in depletion and accumulation with good signal-to-noise. A combination of spatially and spectrally resolved photocurrent measurements is used to separate the intrinsic photoresponse due to SWCNT inter-band absorption from the extrinsic photoresponse due to photogating by the Si substrate. Furthermore, time-resolved photocurrent measurements enable differentiation of the extrinsic displacement photocurrent from the SWCNT intrinsic photoresponse.

For this study, we chose arc-discharge derived 99% semiconducting SWCNTs (P2, Carbon Solutions, Inc.) with a first excitation sub-band ($S_{11}$~1800 nm) below the Si absorption band (~1100 nm).[12,26] SWCNT TFTs were fabricated on 4-layers of zirconia-based self-assembled nanodielectric (Z-SAND)[25] grown on Si substrates using photolithography, reactive-ion etching and lift-off methods. Bottom-contact SWCNT TFTs were then fabricated by transfer of vacuum-filtered semiconducting SWCNT thin films onto the source-drain electrodes (2 nm Ti, 70 nm Au), as described in reference 26. The areal density of SWCNTs was determined to be approximately 17.3 SWCNTs/$\mu m^2$ (See Supporting S1 for details). Both the channel length (L) and the channel width (W) of the devices were varied over the range 5 – 50 μm. To enable wire-bonding on ultra-thin gate dielectrics (~16 nm) with unhindered access to the channel by a 100X objective (NA = 0.95), source-drain electrodes were connected to large metallic pads (500 μm x 2000 µm) on 250 - 500 nm thick e-beam evaporated $SiO_2$ pads ~3 mm away from the SWCNT channel (See Supporting S2 for optical images). The experimental set-up of the SPCM and the SWCNT TFT geometry are shown in Figure 1a. A chopper combined with either a lock-in amplifier or an oscilloscope was used to obtain the spatially and temporally resolved photocurrent measurements, respectively. The photon energy was varied from the Si bandgap (1100 nm) to the SWCNT $S_{11}$ band (1800 nm) by using a tunable coherent white light source (NKT Photonics).

Figure 1b shows the time-dependence of the negative photocurrent of a SWCNT TFT (L = 5 µm, W = 10 µm) in moderate accumulation ($V_g$ = 0 V, $V_d$ = 0.5 V) at excitation wavelengths of 1100 nm, 1200 nm, and 1300 nm. The temporal photoresponse is qualitatively similar across these wavelengths: on a time scale of hundreds of microseconds, the photocurrent reaches a steady-state value that is proportional to the drain bias (the rise time of the preamplifier is 30 microseconds). Figure 1b also shows qualitatively that the magnitude of the steady-state negative photocurrent decreases with longer excitation wavelengths. A comparison between the magnitude of the negative photocurrent and the absorption spectrum of silicon is shown on a log scale in Figure 1c. While the silicon absorption drops sharply around 1200 nm, the negative photoresponse persists to significantly longer wavelengths, which is evidence that some of the observed negative photocurrent is due to absorption by the SWCNTs. The corresponding spatial photocurrent maps are presented in Figures 1d, 1e, and 1f. The above-gap (Fig. 1d) and near-gap (Fig. 1e) excitation induces a negative photocurrent even when the illumination is outside of the SWCNT channel, unlike the sub-gap excitation (Fig. 1f), which only induces a negative photocurrent when the SWCNT channel itself is being illuminated. Additionally, greater contrast is observed within the SWCNT channel for sub-gap excitation, which is likely due to small variations in density within the SWCNT film. This series of images reveals that illumination of the gate plays an important role in the negative photocurrent and that the intrinsic photocurrent signal originating from the excitation of the SWCNTs can be isolated by using photon energies below the Si bandgap. We first discuss the photoresponse arising from illumination at energies above the Si bandgap.

The presence of a negative photocurrent due to above-gap excitation outside the SWCNT channel can be explained by a photogating effect in which excess carriers generated in the n-type Si substrate result in hole accumulation at the Si/Z-SAND interface, inducing band bending in the Si (Fig. 2a).[16] The SWCNTs, which are p-doped in ambient due to atmospheric adsorbates[27,28] are therefore depleted. The dependencies of the negative photocurrent on excitation position, wavelength, and gate voltage are all consistent with photogating. First, the spatial extension of the negative photocurrent away from the SWCNT channel when illuminating above the Si bandgap (1100 nm, Fig. 1d) is established by the large minority carrier (holes) diffusion length (10 – 100 μm) in n-doped Si.[16] Second, as the excitation energy decreases below the Si bandgap (Fig. 1e and 1f), the negative photocurrent away from the SWCNT channel decreases to zero. Third, the transconductance peak corresponds well with the photocurrent peak. Figure 2b (upper plot) shows the device transfer curve ($I_d$ versus $V_g$) as the gate voltage is swept from 2V to -2V with an on/off ratio of ~$10^3$ (see Supporting S2 for output characteristics). The transconductance ($dI_d/dV_g$) is then extracted and plotted with the negative photocurrent as a function of gate voltage in Figure 2c. The correspondence in peak response supports the conclusion that photogating is responsible for the negative photocurrent induced by the above-gap excitation.

The absence of negative photocurrent at large positive $V_g$ (depletion) and large negative $V_g$ (accumulation) can be understood by considering the capacitance-voltage (C-V) curve of a metal-insulator-semiconductor (MIS) capacitor fabricated on 4 layers of Z-SAND (Fig. 2b, lower plot) and band diagrams of the SWCNT/Z-SAND/Si structure (Fig. 2d). The capacitance was measured at 10 kHz, with the Si substrate biased and the metal electrode grounded. The C-V curve shows that the capacitor is in accumulation in region 3 with maximum capacitance ~ 400 F/cm$^2$, and that it transitions into depletion with positive bias applied (region 1). Although there is large band-bending at the Si/Z-SAND interface at $V_g = 2$ V (Fig. 2d, region 1), the light-induced excess holes at the Si/Z-SAND interface do not contribute to the negative photocurrent since the SWCNTs are depleted, resulting in a very small transconductance (Fig. 2c). Instead, as explained below, a positive photocurrent is observed from excitation of the SWCNTs. At $V_g = 0$ V, both band bending at the Si/Z-SAND interface and the transconductance of the device are significant, leading to photogating having a large impact on the photocurrent. At $V_g = -2$ V, there is still a moderate transconductance, but an accumulation region has formed in the Si substrate near the interface mitigating the influence of photocarrier generation in the Si.

We note that the negative photocurrent also depends strongly on the history of the gate bias. The hysteresis in Figure 2b indicates the presence of negative trapped charges at the SWCNT/dielectric interface when the gate voltage is swept from +2 V to -2 V.[28-30] Negative trapped charges at the SWCNT/dielectric interface increase the band bending at the dielectric/substrate interface, resulting in more free holes accumulating at the Si/Z-SAND interface, enhancing the photogating effect. This is in qualitative agreement with our observation that the negative photocurrent is approximately 15X larger when the gate is swept downward (data not shown). Devices fabricated on the present zirconia-based SAND exhibit less hysteresis than the devices made on conventional oxide gate dielectrics such as $SiO_2$ and $Al_2O_3$. Recently, SWCNT TFTs on another SAND variant, vapor-deposited VA-SAND, have also shown negligible hysteresis,[26] indicating that further improvements are possible.

The negative photocurrent that is observed with sub-gap excitation (Figure 1(f)) may arise from a local (1) decrease in free carrier concentration, (2) decrease in mobility, or (3) thermally induced current. A decrease in carrier concentration is not expected given the lack of inversion and the sign of the substrate photogating effect, though sub-gap illumination could change the occupancy of interfacial traps, leading to changes in free carrier concentration. We do expect that the carrier mobility is reduced by local heating of the SWCNTs, which do not have strong thermal connections to either each other or to the back-gate; temperature-dependent conductance measurements (not shown) indicate that the conductance (G) decreases as temperature (T) increases, which leads to the previously observed photo-thermal effect under illumination[9]. A heating induced reduction in conductivity is also consistent with a diminishment in signal approaching the metal contacts, which are effective thermal sinks. The rapid response time suggests that laser induced desorption of adsorbates is not a likely explanation[8,16], and the sign of the photocurrent rules out the bolometric effect observed in thick films (40 nm – 1 μm) of unsorted SWCNTs suspended in vacuum[4].

Neither the negative photocurrent described above nor the positive photocurrent described below appear to originate from a photothermoelectric effect, in which variations in the Seebeck coefficient in the presence of temperature gradients generate potential differences on either side of the illuminated region[17,18]. Previously, photothermoelectric voltages were observed at the edge of the suspended SWCNT films due to induced temperature gradients[17] and near the metal contacts on supported SWCNT films due to the difference in the Seebeck coefficient between SWCNTs and metal[18]. Thus, the photothermoelectric effect produces complementary positive and negative signals depending on excitation location, where the difference in the Seebeck coefficient and/or induced temperature gradient determines the sign of the photoresponse. In contrast, our devices showed either a negative or a positive photocurrent (as discussed below) depending on the biasing condition, but not both positive and negative photocurrents at a given biasing condition. Thus, a heating induced decrease in conductance (the "photo-thermal effect" in reference 9) is the likely origin of the negative photocurrent under sub-gap illumination for our devices and experimental conditions, and this signal dominates any underlying photothermoelectric response. Additional analyses are provided below in the discussion of the depletion mode photoresponse.

Next, we discuss the mechanisms behind the intrinsic and extrinsic photocurrents generated by illuminating a SWCNT TFT in depletion. The time-dependent positive photocurrent from the same SWCNT device in the depletion region (Fig. 3a, $V_g = 2$ V) exhibits two distinct components: a transient response (green lines) and a steady-state response (black lines). The transient photocurrent response is attributed to an extrinsic displacement photocurrent arising from illumination-induced band-bending and associated changed in the charge density accumulated across the SWCNT/Z-SAND/Si capacitor. As expected, the energy dependence of the accumulated charge (integrated displacement photocurrent) tracks the Si absorption coefficient (Fig. 3b). In addition, this displacement photocurrent is observed everywhere between the electrodes and not just in the SWCNT channel, a clear indication of an extrinsic response.

The second component of the positive photocurrent is an intrinsic, steady-state photocurrent that persists after the laser has been 'on' for several milliseconds and is attributed to the generation of free carriers from inter-band absorption[20] in the SWCNT thin film. Figures 3d and 3e show positive photocurrent maps of the SWCNT TFT at an excitation wavelength of 1300 nm, but at opposite biases. In both cases, the positive photocurrent is stronger at the grounded (reverse biased) electrode. The schematic in Figure 4a describes the band bending at the reverse-biased electrode for a p-type semiconductor, resulting in increased dissociation of optically generated excitons and increased positive photocurrent. Additionally, measurements in depletion ($V_g$ = 2 V) result in a larger built-in electric field at the grounded electrode ($|V_s - V_g|$ = 2 V) than at the biased electrode ($|V_d - V_g|$ = 1.5 V; Fig. 3d and 3e). Further evidence that the positive photocurrent results from free carriers from inter-band absorption is found in spectrally resolved photocurrent measurements. In Figure 4b, the positive persistent photocurrent is compared with the absorption spectrum for 99% semiconducting SWCNT thin films in the range 1200 nm – 1800 nm.[31] Previously, photoluminescence and electroluminescence spectra of similar monodisperse SWCNTs showed a broad peak at ~1950 nm (red-shifted from the first excited state $E_{11}$, 1800 nm) due to excitonic energy transfer to largest diameter SWCNTs in the network.[12,32] The spectral range of the present source was insufficient to probe the entire peak, but the energy dependence is consistent with inter-band excitations in SWCNTs.[33] The linear dependence of photocurrent on light intensity (Supporting Fig. S5) suggests that dissociation of optically generated excitons is rapid, and rules out the nonlinear thermal effects such as Benedicks effect[34].

Finally, we consider again the possibility of a photothermoelectric origin for the persistent positive photocurrent. The decay time of the transient photothermoelectric current in supported SWCNT films was found to be 0.3 – 2.2 s.[18] In contrast, the rise time of the persistent photocurrent for our supported films was faster than 30 μs, the preamplifier limited risetime. In addition, the persistent photocurrent increased monotonically with drain bias (Supporting Fig.

S6), which is not expected for a photothermoelectric response, but is expected for drift current of free carriers resulting from inter-band excitations. [8]

In conclusion, scanning photocurrent microscopy and spectroscopy studies of 99% pure semiconducting SWCNT TFTs were carried out for devices based on hybrid organic-inorganic gate dielectrics. The combination of monodisperse SWCNTs and high capacitance (~400 nF/cm$^2$) gate dielectrics enables TFTs with low bias device operation (4 V) and channel conductance modulation over 3-4 orders of magnitude. Simultaneous measurements of spatially, spectrally, and temporally resolved photocurrent were utilized to distinguish between the extrinsic photocurrent and intrinsic photoresponses of monodisperse SWCNTs. Substrate-induced photogating contributed to significant negative photocurrent in the accumulation regime. In contrast, photocurrent in the depletion region arises from an extrinsic transient displacement current and intrinsic photocurrent from inter-band excitation in the SWCNTs.

**Acknowledgement:**

This research was supported by the National Science Foundation (DMR-1006391 and DMR-1121262) and by the Nanoelectronics Research Initiative at the Materials Research Center of Northwestern University. HNA acknowledges support from a NASA Space Technology Research Fellowship. We thank Drs. R. Divan and L. Ocala of the Center for Nanoscale Materials, Argonne National Laboratory, Dr. N. Basit of the Micro/Nano Fabrication Facility, Northwestern University, and R. Lajos, Nanotechnology Core Facility, University of Illinois, Chicago for assistance with clean room fabrication. We also thank Dr. J. Hyun for helpful discussions.

**Figures:**

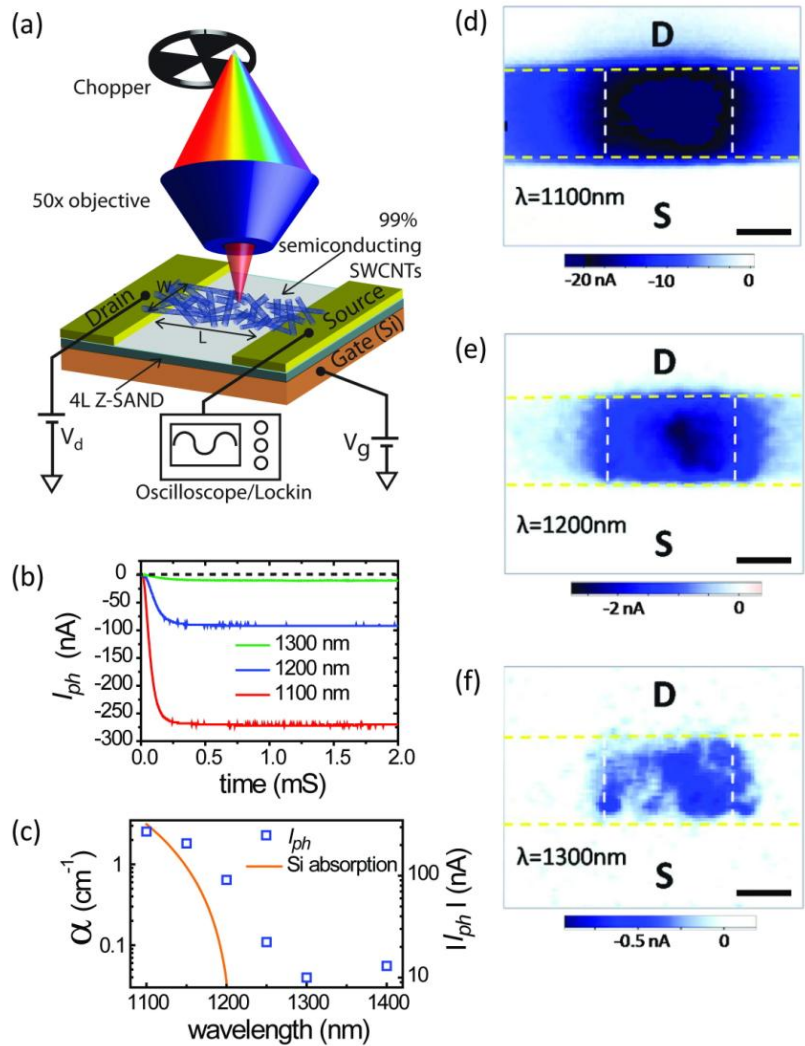

Figure 1. a) Schematic of the experimental set-up for scanning photocurrent imaging of single-walled carbon nanotube (SWCNT) thin film transistors (TFTs) on 4 layers of the self-assembled nanodielectric Z-SAND. b) Negative photocurrent response measured by oscilloscope at excitation wavelengths of 1100 nm, 1200 nm and 1300 nm. c) Magnitude of the negative photocurrent plotted as a function of excitation wavelength for comparison with the absorption coefficient of Si. d), e), f) Spatial maps of negative photocurrent at wavelengths 1100 nm, 1200 nm and 1300 nm, respectively. The dotted lines indicate the SWCNT thin film channel. The scale bars are 3 μm.

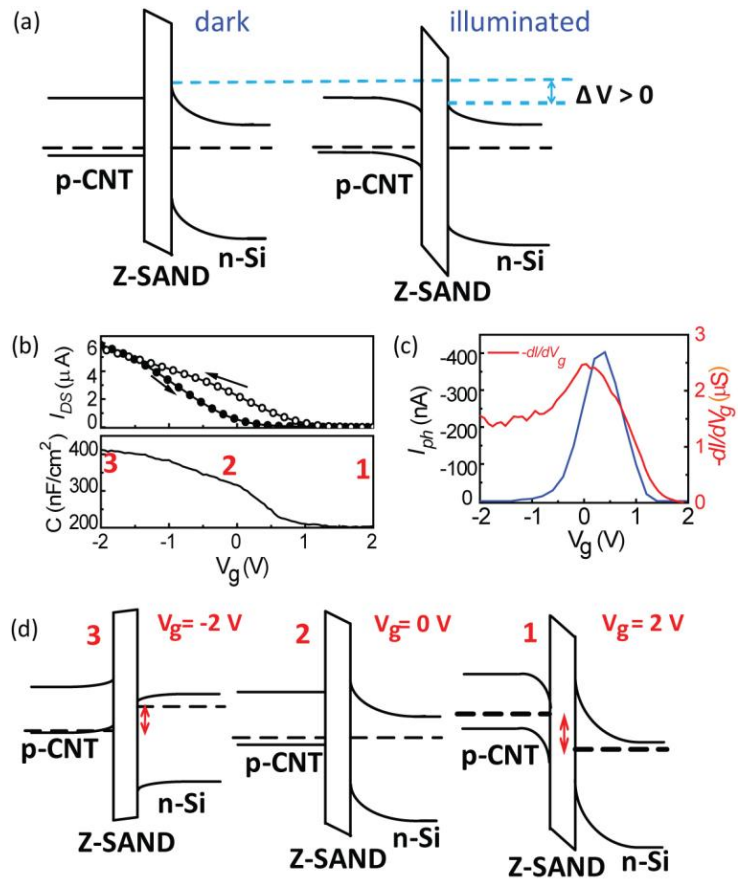

Figure 2. a) Band diagrams of the SWCNT/Z-SAND/Si structure in the dark and under illumination showing light-induced band-bending in the Si substrate. Negative photocurrent is induced by photogating from excess carriers at the Si/Z-SAND interface. b) Transfer characteristics (upper) of a SWCNT TFT with L = 5 μm and W = 10 μm are compared with capacitance-voltage curve (bottom) of MIS capacitors on 4L Z-SAND. c) Negative photocurrent and transconductance are plotted as a function of the gate voltage. d) Evolution of the band diagram of the SWCNT/Z-SAND/Si structure as the gate voltage is swept from 2 V to -2 V.

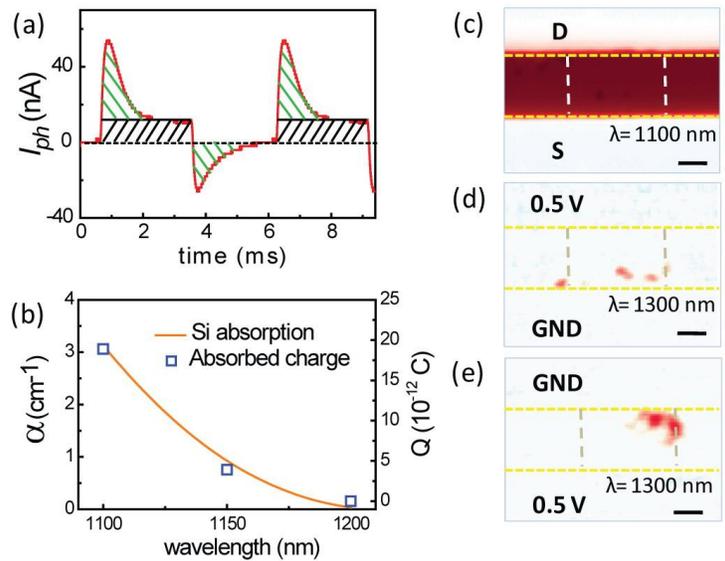

Figure 3. a) Time dependent positive photocurrent response showing transient displacement current (green lines) as well as a persistent photocurrent (black lines) at an excitation 1100 nm. b) Net displacement charge is plotted as a function of wavelength for comparison with the Si absorption coefficient. c) Spatial map of the transient positive photocurrent at an excitation wavelength of 1100 nm. The scale is from 0 to 5 nA. d), e) Spatial maps of persistent positive photocurrent at an excitation wavelength of 1300 nm when the bottom and top electrodes are grounded, respectively. The scale in d) and e) is from 0 to 35 pA and 0 to 50 pA, respectively. All scale bars are 3 μm.

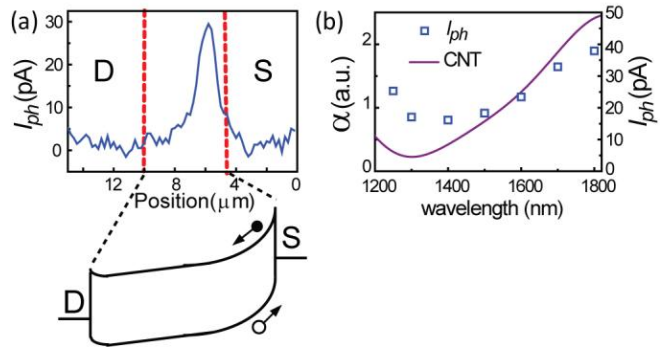

Figure 4.a) Width-integrated positive photocurrent plotted as a function of channel length. A schematic band diagram suggesting that positive photocurrent near the grounded electrode arises from free carriers generated by inter-band excitation of the SWCNTs. b) Positive photocurrent plotted as a function of wavelength and compared with the absorption spectrum of a 99% pure semiconducting SWCNT thin-film. (See Ref. 31)[34]

**References:**

See supplementary material at [URL will be inserted by AIP] for characterization of density of SWCNT network, electrical characterization of SWCNT TFTs, and illumination power dependence of the intrinsic photocurrent.